\documentclass[reprint,showpacs,amsmath,amssymb,pra]{revtex4-1}
\usepackage{graphicx}
\usepackage{dcolumn}
\usepackage{bm}

\begin{document}

\title{Compression and collisions of chirped pulses in a dense two-level medium}

\author{Denis V. Novitsky}
\email{dvnovitsky@gmail.com} \affiliation{B. I. Stepanov Institute
of Physics, National Academy of Sciences of Belarus, Nezavisimosti
Avenue 68, BY-220072 Minsk, Belarus}

\date{\today}

\begin{abstract}
Using numerical simulations, we study propagation of
linearly-chirped optical pulses in a homogeneously broadened
two-level medium. We pay attention to the three main topics --
validity of the rotating-wave approximation (RWA), pulse
compression, and collisions of counter-propagating pulses. The cases
of long and single-cycle pulses are considered and compared with
each other. We show that the RWA does not give a correct description
of chirped pulse interaction with the medium. The compression of the
chirp-free single-cycle pulse is stronger than of the chirped one,
while the opposite is true for long pulses. We demonstrate that the
influence of chirp on the collisions of the long pulses allows to
control the state of the transmitted radiation: the transmission of
the chirp-free pulse can be dramatically changed under collision
with the chirped counter-propagating one, in sharp contrast to the
case when both pulses are chirped. On the other hand, the collisions
of the chirped single-cycle pulses can be used for precise control
of medium excitation in a narrow spatial region.
\end{abstract}


\maketitle

\section{Introduction}

Since the discovery of self-induced transparency (SIT)
\cite{McCall1, McCall2}, light interaction with two-level quantum
media attracted much attention and was discussed in a number of
books \cite{Allen, MaimistovBook} and review papers \cite{Kryukov,
Lamb, Poluektov, Maimistov}. This interest is, to a great extent,
due to the fundamental importance of the semiclassical two-level
model which is one of the basic models of nonlinear optics and laser
physics. Besides the SIT itself, there was a deep investigation of
other nonlinear effects in two-level media, such as intrinsic
(mirrorless) optical bistability \cite{Bonifacio, Hopf}, influence
of local-field correction (near dipole-dipole interactions) on SIT
solitons and optical switching \cite{Bowd93, Cren92, Bowd91},
population control with specially constructed pulses \cite{Golubev},
incoherent soliton generation \cite{Afan02}, collisions of solitons
\cite{Shaw, Novit2011}, solitons in periodically modulated two-level
media \cite{Kozhekin, Kurizki, Lemeza}, etc. A more recent topic is
connected with study of few-cycle and sub-cycle pulses in the
two-level media when the standard rotating-wave approximation (RWA)
turns out to be invalid \cite{Ziolkowski, Novit2012b, Cai2013}; see
also the recent reviews \cite{Leblond, Frantzeskakis} and references
therein.

Additional degree of freedom is provided by chirp, i.e. temporal
variation of the carrier frequency of the pulse. Influence of chirp
on pulse propagation in the two-level medium is under examination,
at least, from the 1980s \cite{Manassah}. More recent theoretical
studies allowed to find the analytical solutions for a certain class
of chirped pulses \cite{Jha} and investigate the validity of the RWA
for the so-called ultrachirped pulses \cite{Ibanez}, showed the
splitting of chirped pulses with particular spectral composition
\cite{Yao}, demonstrated the soliton formation from a two-component
chirped pulse \cite{Xu} and the coherent control of spectral shifts
\cite{Song}, analyzed excitation of the medium with the sub-cycle
and single-cycle chirped pulses and formation of sub-cycle solitons
\cite{Astapenko, Cai2013, Cai2015}, etc.

In this paper, we consider some aspects of medium-light interaction
in the case of linearly-chirped short pulses. We focus on the three
main questions which, as far as we know, were not studied in detail
previously -- test of the RWA validity, pulse compression and
collisions of counter-propagating pulses. We use numerical
simulation technique described briefly in Section \ref{eqpars} to
directly verify the validity of the RWA for description of
linearly-chirped pulse propagation and to study the compression of
such pulses and soliton formation. In Section \ref{long}, the pulses
are suggested to be long enough, so that the RWA violation cannot be
connected with the processes on the single-cycle scale studied
previously \cite{Novit2012b}. As to collisions of
counter-propagating chirped pulses inside the medium, the present
study continues our previous work where interaction of chirp-free
pulses in both homogeneously and inhomogeneously broadened two-level
media was analyzed \cite{Novit2011, Novit2012a, Novit2014}. It was
shown that, changing intensity of the first pulse, one can
effectively control the transmission of the second pulse. Here we
study the influence of chirp on collisions of counter-propagating
pulses and the possibility to control the parameters of transmitted
radiation with the chirped pulses. In Section \ref{cycle}, the case
of single-cycle pulses is considered and compared with the results
obtained for the long pulses. The paper is completed with the brief
Conclusion.

\section{\label{eqpars}Main equations and parameters}

We describe light propagation in the homogeneously broadened
two-level medium beyond the RWA and the slowly-varying envelope
approximation (SVEA) with the Maxwell--Bloch equations as given in
our previous publication \cite{Novit2012a}:
\begin{eqnarray}
\frac{\partial^2 \Omega}{\partial \xi^2}&-& \frac{\partial^2
\Omega}{\partial \tau^2}-2 i \frac{\partial \Omega}{\partial \xi}-2
i \frac{\partial \Omega}{\partial
\tau} \nonumber \\
&&=6 \epsilon \left(\frac{\partial^2 p}{\partial \tau^2}+2 i
\frac{\partial p}{\partial \tau}-p\right), \label{Maxdl}
\end{eqnarray}
\begin{eqnarray}
\frac{d p}{d \tau} &=& i \delta p + \frac{i}{2} (\Omega + s \Omega^*
e^{-2 i (\tau-\xi)}) w - \gamma'_2 p, \label{polardl}
\end{eqnarray}
\begin{eqnarray}
\frac{d w}{d \tau} &=& i (\Omega^* p - \Omega p^*) + i s
\left(\Omega p
e^{2 i (\tau-\xi)} - \Omega^* p^* e^{-2 i (\tau-\xi)} \right) \nonumber \\
&&- \gamma'_1 (w+1), \label{inversdl}
\end{eqnarray}
where $\tau=\omega t$ and $\xi=kz$ are the dimensionless time and
distance; $\Omega=(\mu/\hbar \omega) A$ is the dimensionless field
amplitude (normalized Rabi frequency); $A$ and $p$ are the complex
amplitudes of the electric field and atomic polarization,
respectively; $w$ is the inversion (difference between populations
of excited and ground states); $\delta=\Delta
\omega/\omega=(\omega_0-\omega)/\omega$ is the normalized frequency
detuning; $\omega_0$ is the frequency of atomic resonance; $\omega$
is the light carrier frequency; $\mu$ is the dipole moment of the
quantum transition; $\gamma'_{1,2}=\gamma_{1,2}/\omega$ are the
normalized relaxation rates  of population and polarization,
respectively; $\epsilon= \omega_L / \omega=4 \pi \mu^2 C/3 \hbar
\omega$ is the dimensionless parameter of interaction between light
and matter (normalized Lorentz frequency); $C$ is the concentration
(density) of two-level atoms; $k=\omega/c$ is the wavenumber; $c$ is
the speed of light, and $\hbar$ is the Planck constant. Asterisk
stands for complex conjugation. We introduced here the auxiliary
two-valued coefficient $s$, so that $s=0$ corresponds to the RWA
(absence of ``rapidly rotating'' terms), while $s=1$ is used in the
general case.

Further, we solve Eqs. (\ref{Maxdl})--(\ref{inversdl}) numerically
choosing the appropriate value of $s$. The numerical approach is the
same as in our previous publication \cite{Novit2012a}, more details
on it can be found in \cite{Novit2009}. We perform calculations for
the following parameters of the medium and light: the relaxation
rates $\gamma_1=1$ and $\gamma_2=10$ ns$^{-1}$ are large enough, so
that we are in the regime of coherent light-matter interaction; the
detuning $\delta=0$ (exact resonance, $\omega=\omega_0$); the
central light wavelength $\lambda=2 \pi c / \omega_0=0.83$ $\mu$m;
and the strength of light-matter coupling $\omega_L=10^{11}$ s$^{-1}
\ll \omega$. For this choice of parameters, the inequality $\Omega
\omega \gg \omega_L$ is valid, so that we can neglect here the
so-called local field effects \cite{Novit2010}. The medium is
supposed to be initially in the ground state ($w=-1$).

In this paper, we consider the pulses of Gaussian shape with linear
chirp, so that for the incident normalized Rabi frequency (electric
field amplitude) we have $\Omega=\Omega_p \exp(-(t-t_0)^2/2 t^2_p +
i \beta \omega_0^2 t^2)$, where $\beta$ is the dimensionless chirp
parameter (chirp normalized by $\omega_0^2$), i.e. the instantaneous
carrier frequency changes linearly with time as $\omega_i
(t)=\omega_0 + 2 \beta \omega_0^2 t$. The duration of the pulse
$t_p$ is defined through the number of cycles $N$ as $t_p=N T/2
\sqrt{\ln 2}$, where $T=\lambda/c$ is the period of electric field
oscillations. The parameter $t_0$ governs the instant of maximum of
the pulse intensity (the peak offset). It is important to note that
the instantaneous frequency is not equal to $\omega_0$ at the pulse
peak: $\omega_i (t)$ grows linearly from $\omega_0$ at $t=0$ and, at
$t=t_0$, differs from this initial frequency more or less
significantly. The peak Rabi frequency $\Omega_p$ is measured in the
units of $\Omega_0=\lambda/\sqrt{2 \pi} c t_p$ corresponding to the
chirp-free pulse area $2 \pi$.

\section{\label{long}Long pulses}

\begin{figure}[t!]
\includegraphics[scale=0.9, clip=]{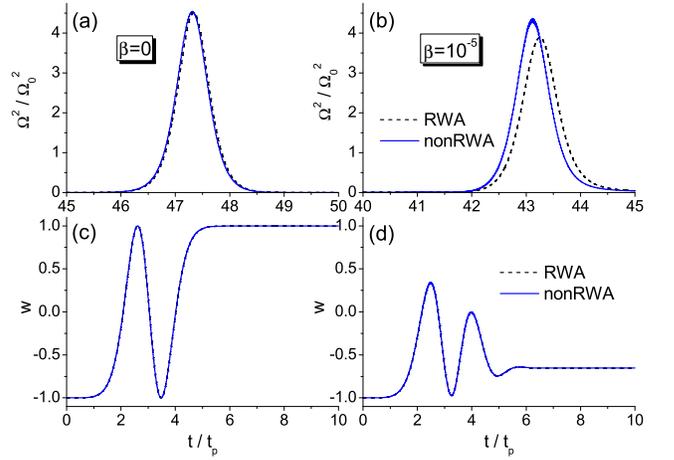}
\caption{\label{fig1} (Color online) (a, b) The profiles of
transmitted radiation and (c, d) inversion dynamics (at the medium
entrance) for the incident pulse (a, c) without chirp and (b, d)
with the chirp $\beta=10^{-5}$. The layer thickness is $L=1000
\lambda$, the pulse amplitude is $\Omega_p=1.5 \Omega_0$.}
\end{figure}

\begin{figure}[t!]
\includegraphics[scale=0.9, clip=]{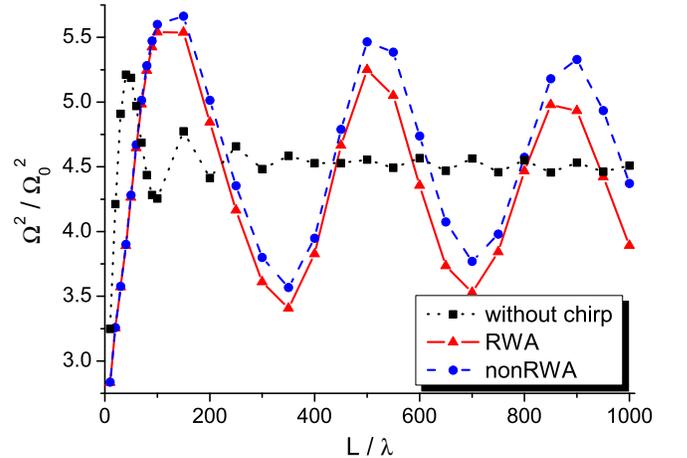}
\caption{\label{fig2} (Color online) The dependence of the peak
intensity of the transmitted pulse calculated for the chirp values
of $\beta=0$ and $\beta=10^{-5}$. In the latter case, both RWA and
non-RWA calculations were performed. The pulse amplitude is
$\Omega_p=1.5 \Omega_0$.}
\end{figure}

In this section, we consider the long pulses with the number of
cycles $N=50$ and the peak offset $t_0=3 t_p$. We start with the
dynamics of a single chirped pulse in the two-level medium focusing
on pulse transmission through layers of different thicknesses. One
of our main intentions is to test the applicability of the RWA for
description of pulse dynamics. Figure \ref{fig1} shows the
comparison of the intensity profiles and inversions calculated with
and without the RWA for the chirp-free pulse and for the pulse with
the chirp parameter as large as $\beta=10^{-5}$. The pulse amplitude
$\Omega_p=1.5 \Omega_0$ corresponds to the area $3 \pi$ (in the
chirp-free limit), so that the non-chirped pulse leaves the medium
in the fully inverted state [see Fig. \ref{fig1}(c)]. This is not
the case for the chirped pulse [Fig. \ref{fig1}(d)]: the medium
stays only partially excited, perhaps, because of violation of the
resonance conditions due to sweeping of the pulse frequency. It is
seen that calculations with the RWA ($s=0$) and without it ($s=1$)
give essentially the same dynamics of the inversion, but not of the
intensity profiles of the transmitted radiation. The RWA works
perfectly in the case of chirp-free pulse [Fig. \ref{fig1}(a)], but
the results for chirped pulse diverge [Fig. \ref{fig1}(b)]. This
difference between the RWA and non-RWA profiles seems to be the
result of small deviations from the exact dynamics which accumulate
as the pulse propagates in the medium.

This conclusion is corroborated in Fig. \ref{fig2} where the
dependence of the transmitted pulse peak intensity on the medium
thickness $L$ is shown. It is clearly seen that the difference
between the RWA and non-RWA curves calculated for the pulses with
$\beta=10^{-5}$ grows with $L$. Calculations using the RWA give
lower intensities in comparison with the general model. It is also
interesting to compare compression of the pulses with and without
linear chirp which can be traced by change of the peak intensity of
the pulse. It is seen in Fig. \ref{fig2} that, for the chirp-free
pulse (black squares), the peak intensity rises from
$\Omega_p^2=2.25 \Omega_0^2$ to the maximum of about $5.25
\Omega_0^2$ at the comparatively short distance in the medium (less
than $100 \lambda$) and then, after some oscillations, tends to the
quasi-stationary level of about $4.5 \Omega_0^2$ (formation of the
$2 \pi$ SIT soliton). These oscillations are much more pronounced
for the chirped pulse: though the average level of compression is
essentially the same (peak with $4.5 \Omega_0^2$), the scatter of
data around this mean value allow to obtain at different medium
thicknesses strongly differing results (from $3.5 \Omega_0^2$ to
$5.5 \Omega_0^2$). These oscillations relax and, perhaps, result in
(quasi)solitonic pulse formation, but much more slowly than in the
case of $\beta=0$. All in all, the chirped pulses may be useful to
obtain stronger compressions, especially at larger medium
thicknesses than the chirp-free pulses.

\begin{figure}[t!]
\includegraphics[scale=0.9, clip=]{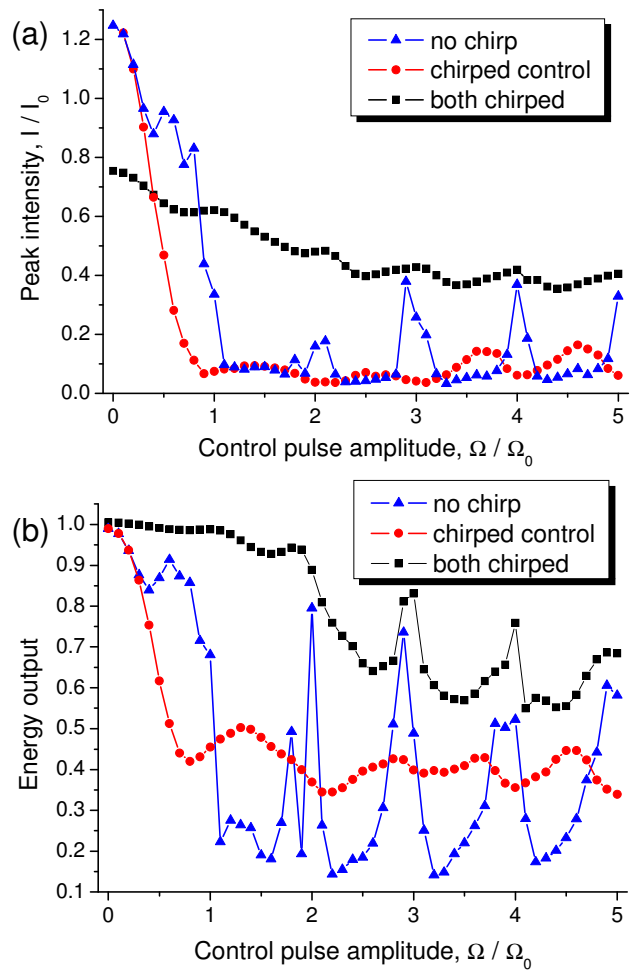}
\caption{\label{fig3} (Color online) The dependencies of (a) the
peak intensity of the signal pulse and (b) the part of its energy at
the output on the amplitude of the control (counter-propagating)
pulse. Calculations were performed for three cases: (i) both pulses
without chirp, (ii) the control pulse is chirped, but the signal one
is not, (iii) both pulses are chirped. The chirp is $\beta=10^{-5}$,
the medium thickness $L=350 \lambda$, the signal pulse amplitude
$\Omega_p=\Omega_0$.}
\end{figure}

\begin{figure}[t!]
\includegraphics[scale=0.9, clip=]{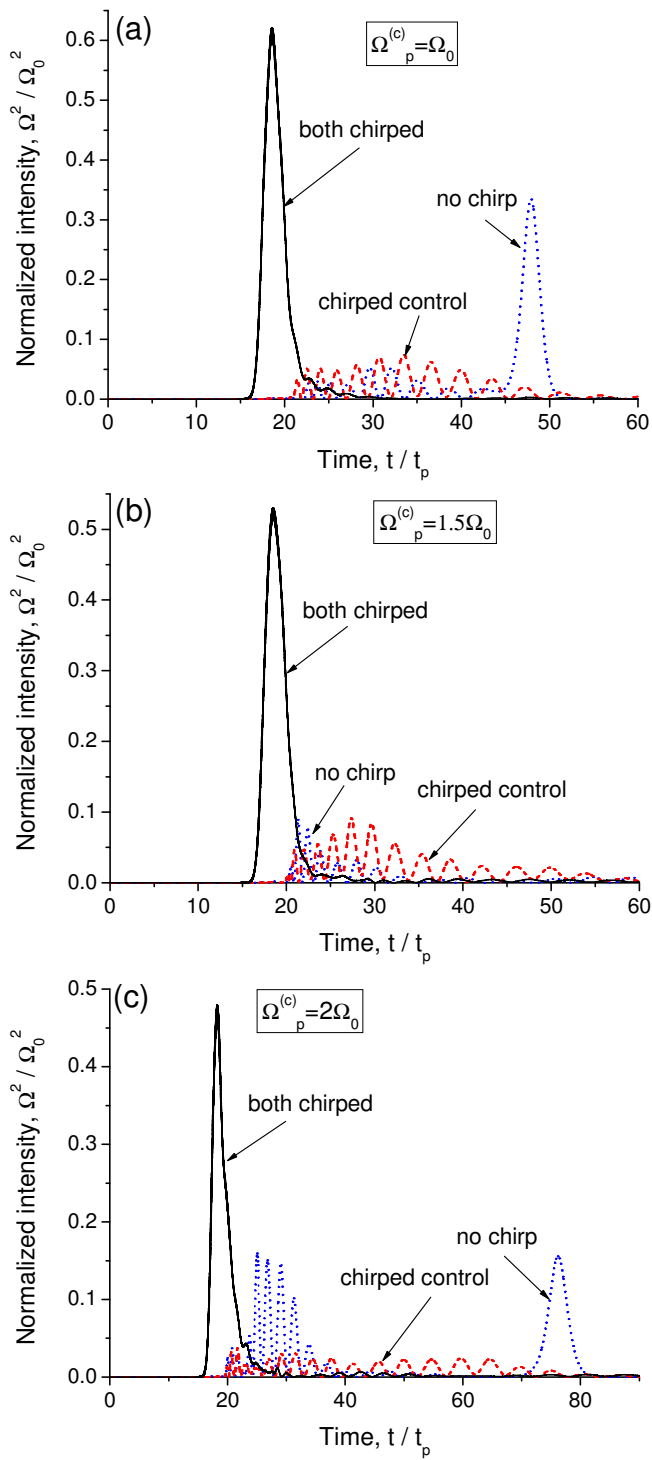}
\caption{\label{fig4} (Color online) The profiles of the signal
pulse after interaction with the control (counter-propagating) one.
The latter's amplitude $\Omega^{(c)}_p$ is (a) $\Omega_0$, (b) $1.5
\Omega_0$, (c) $2 \Omega_0$. Calculations were performed for the
same three cases and the same parameters depicted in Fig.
\ref{fig3}.}
\end{figure}

Let us consider the situation of colliding counter-propagating
pulses and study the influence of chirp on their interaction. We
call one of the pulses (forward propagating) the signal pulse, while
the counter-propagating one is named the control pulse. We focus on
the following question: How the control pulse influences the
properties of the signal one? To find the answer, we fix the
amplitude of the signal pulse to $\Omega^{(s)}_p=\Omega_0$ ($2 \pi$
pulse) and change the amplitude of the control pulse
$\Omega^{(c)}_p$ from $0$ to $5 \Omega_0$ (area from $0$ to $10
\pi$). The results of calculations are shown in Fig. \ref{fig3}. We
consider three cases: (i) both pulses have no chirp, (ii) the
control pulse is chirped, but the signal one is not, (iii) both
pulses are chirped.

The starting point is the interaction of chirp-free pulses which was
studied in detail in our previous works \cite{Novit2011,
Novit2012b}. The curves for the signal peak intensity and the energy
output (the part of signal pulse energy left the medium after the
time $100 t_p$) as a function of control pulse amplitude in this
chirp-free case are shown with blue triangles in Fig. \ref{fig3}.
These curves demonstrate the pronounced periodicity: they have
maxima at the control pulse amplitudes $n \Omega_0$, where $n$ are
the integer numbers; this condition corresponds to the areas of the
control pulse $2 \pi n$. At these maxima, there are both the
high-intensity signal soliton and low-intensity oscillations
(precursor) at the output of the medium as shown in Fig.
\ref{fig4}(a) and \ref{fig4}(c) (see blue dotted lines). On the
contrary, at the minimal peak intensity and output energy (when $n$
is half-integer), we have only precursor at the output
[\ref{fig4}(b)], so that most part of signal energy (more than $80
\%$) is stored inside the medium and leaves it slowly in the form of
fluorescent radiation on the time scales of relaxation times. Thus,
changing control pulse amplitude between integer and half-integer
numbers of $\Omega_0$ allows to switch on and off the solitonic
component of the signal pulse at the medium output.

Now, let us consider the case when both signal and control pulses
are chirped with $\beta=10^{-5}$ (black squares in Fig. \ref{fig3}).
In this case, the periodicity is much less pronounced, while the
peak intensity and the output energy of the signal pulse remain
relatively large at every value of the control pulse amplitude. This
means that the high-intensity signal pulse only slowly changes its
peak when we take different amplitudes of the control pulse as can
be corroborated directly in Fig. \ref{fig4} (black dashed lines).
Thus, chirped pulses seem to be not suitable to effectively control
one another.

The last case which we should consider is the chirp-free signal
pulse interacting with the chirped control one (red circles in Fig.
\ref{fig3}). The calculations in this case give an interesting and
unexpected result: the peak intensity of the signal pulse drops to
the very low values and does not manifest such sharp periodic peaks
as in the case of chirp-free control pulse. At the same time, the
output energy remains relatively large at all values of control
amplitude. As Fig. \ref{fig4} (red solid curves) shows, we have only
precursor oscillations and no soliton regardless of the control
pulse amplitude. This is true already for $\Omega^{(c)}_p=\Omega_0$
which means that we can use chirped control of lower intensity to
get rid of chirp-free soliton and save only low-intensity
oscillations. Thus, chirped control pulse appears to be very
effective mean to destroy the chirp-free signal pulse.

\section{\label{cycle}Single-cycle pulses}

\begin{figure}[t!]
\includegraphics[scale=0.9, clip=]{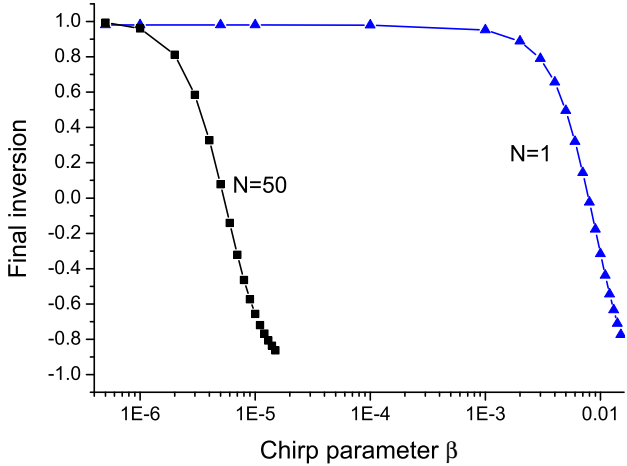}
\caption{\label{fig5} (Color online) The dependence of the final
state of inversion on the chirp parameter for long and single-cycle
pulses. The pulse amplitude is $\Omega_p=1.5 \Omega_0$.}
\end{figure}

\begin{figure}[t!]
\includegraphics[scale=0.9, clip=]{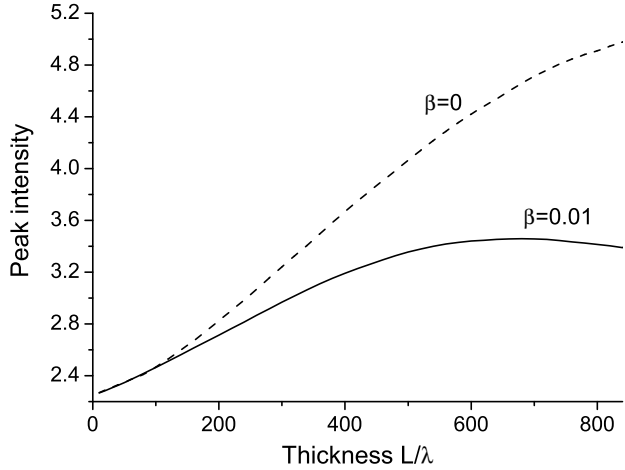}
\caption{\label{fig6} The peak intensity of the chirped and
non-chirped single-cycle pulses transmitted through the medium of
different thicknesses.}
\end{figure}

\begin{figure}[t!]
\includegraphics[scale=0.9, clip=]{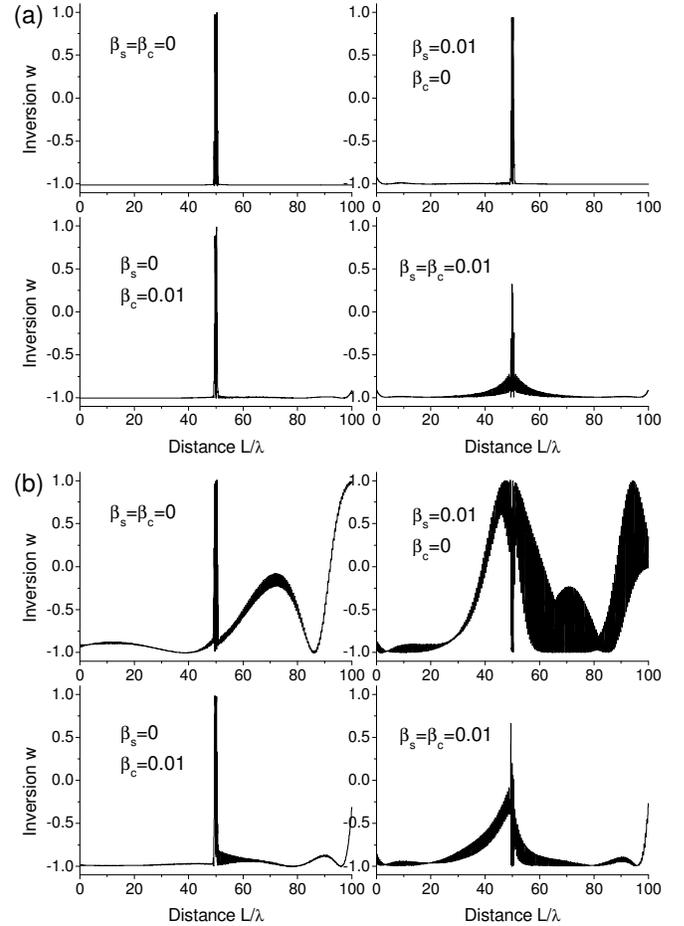}
\caption{\label{fig7} Distributions of inversion along the layer of
two-level medium after collision of signal and control pulses with
different chirps and with amplitudes as follows: (a)
$\Omega^{(s)}_p=\Omega^{(c)}_p=\Omega_0$, (b)
$\Omega^{(s)}_p=\Omega_0$, $\Omega^{(c)}_p=1.5\Omega_0$.}
\end{figure}

In this section, we turn to another, fundamentally different case of
the pulse containing only one cycle of electromagnetic oscillations
($N=1$). As previously, the pulse amplitude is $\Omega_p=1.5
\Omega_0$. The peak offset $t_0=5 t_p$ is taken larger than in the
previous section for the chirp to have more effect after only a
single cycle. First of all, let us illustrate the importance of
pulse duration for pronounced chirp influence. Figure \ref{fig5}
shows the dependence of the final state of inversion (the
steady-state inversion established in the medium after pulse
passage) on the chirp parameter $\beta$. At low chirps, we see the
fully inverted state as expected in the case of $3 \pi$ pulse; the
final state of inversion gradually lowers as the chirp grows. One
could think that the chirp effect on pulse propagation is due to the
frequency shift and then make a simple estimate of the chirp
parameter $\beta_2$ needed for the pulse of $N_2$ cycles to give the
same effect as a pulse with $N_1$ cycles and chirp $\beta_1$:
\begin{equation}
\beta_2=\beta_1 \frac{N_1}{N_2}. \label{estim}
\end{equation}
Taking $N_1=50$, $N_2=1$, and $\beta_1=10^{-5}$, we have $\beta_2=5
\cdot 10^{-4}$. Comparison with the data from Fig. \ref{fig5} shows
that Eq. (\ref{estim}) strongly underestimates the needed chirp: the
final state of inversion (approximately $-0.7$) reached for the long
pulse with parameters listed above is realized for the single-cycle
pulse only at $\beta_2 \approx 10^{-2}$. Thus, in order to have
strong chirp effect, we consider in this section the single-cycle
pulse with the parameter as large as $\beta=0.01$.

Next question to be discussed is the problem of pulse compression.
According to Fig. \ref{fig2}, the long chirped pulse after
transmission through the medium has larger peak intensity, i.e. it
can be compressed stronger than the chirp-free one. What about
single-cycle pulse? Figure \ref{fig6} shows the dependence of the
peak intensity on the medium thickness for such ultra-short pulses.
It is seen that, contrary to the situation considered in the
previous section, compression of the chirped single-cycle pulse is
less effective in comparison with the chirp-free one. The reason for
this is not clear. We can speculate that it may be somehow connected
with the mechanism of soliton formation which can differ depending
on the pulse duration.

Finally, we should analyze the situation of colliding single-cycle
pulses. As previously, we call the forward and backward propagating
pulses ``signal'' and ``control'' ones. As opposed to the case of
long pulses, the collision of single-cycle ones does not result in
any significant energy losses because of very wide spectrum of such
pulses and very narrow region of their collision. Therefore, the
pulses propagate almost uninfluenced after collision and there is no
possibility to control their intensity and other properties.
However, one can use the collisions of single-cycle pulses as a
means to control the state of the medium. Figure \ref{fig7} shows an
example of such control. Contrary to the collisions of long pulses
where medium excitation can be reached in a certain wide spatial
range, here the medium is excited in much more small volume and can
be obtained using thinner layers. Consider the case of two pulses
with the identical amplitudes $\Omega_0$ [Fig. \ref{fig7}(a)]. As a
result of collision of two chirp-free pulses ($\beta_s=\beta_c=0$),
the medium is fully inverted in a very narrow spatial interval near
the center of the layer. Introducing chirp into signal or control
pulse leaves the result almost unchanged. Only when both pulses are
chirped ($\beta_s=\beta_c=0.01$), the level of inversion becomes
significantly lower. Obviously, changing the value of the chirp
parameter, one can obtain the needed level of inversion. Temporal
delay of launching of one of the pulses allows to shift the
excitation region as one desires.

For the pulses with different amplitudes, the picture of collision
is much more complex [Fig. \ref{fig7}(b)]. Generally, the
distributions of inversion in this case are asymmetric with respect
to the center of the layer. However, the level of this asymmetry can
be controlled with the chirp. For example, the distribution in the
case of chirped control pulse is almost symmetric, especially
comparing with the case when both pulses are chirp-free. Thus, with
the chirp, we have one more degree of freedom to control the state
of the medium.

\section{\label{concl}Conclusion}

In summary, we have performed numerical simulations of
linearly-chirped pulse propagation in the homogeneously broadened
two-level medium. First, we have considered long (multi-cycle)
pulses. In particular, we have studied transformations of such long
pulses and showed that, generally, the RWA does not give a correct
description of chirped pulse interaction with the medium. We have
also investigated the collisions of the counter-propagating long
pulses and the influence of chirp on their interaction. Our
calculations have demonstrated that, if both pulses are chirped,
there is only weak dependence of the signal pulse transmission on
the control pulse amplitude. On the contrary, parameters of
chirp-free signal can be dramatically changed with the help of the
chirped control pulse. These dependencies can be used to control
radiation in resonant media which seems to be perspective for
all-optical logic and other applications.

Second, we have performed calculations with the single-cycle pulses
and compared with the case of long pulses. We have shown that
compression of the chirp-free single-cycle pulse is stronger than of
the chirped one, while the opposite is true for the long pulses. As
to collisions, the interaction of the single-cycle pulses can be
used to control the state of the medium in the given spatial region
but not the state of the pulses themselves. The chirp gives the
additional possibility to change the spatial distribution of medium
inversion which can be used for the all-optical precise control of
medium excitation level.


\begin{thebibliography}{0}
\bibitem{McCall1} S. L. McCall and E. L. Hahn, {\prl} {\bf18}, 908 (1967).
\bibitem{McCall2} S. L. McCall and E. L. Hahn, {Phys. Rev.} {\bf183}, 457 (1969).
\bibitem{Allen} L. Allen and J.H. Eberly, \textit{Optical Resonance and Two-Level Atoms} (Wiley, New York, 1975).
\bibitem{MaimistovBook} A. I. Maimistov and A. M. Basharov, \textit{Nonlinear Optical Waves} (Kluwer Academic Publishers, Dordrecht, 1999).
\bibitem{Kryukov} P.G. Kryukov and V.S. Letokhov, {Sov. Phys. Usp.} {\bf 12}, 641 (1970).
\bibitem{Lamb} G. L. Lamb Jr., {\rmp} {\bf43}, 99 (1971).
\bibitem{Poluektov} I. A. Poluektov, Yu. M. Popov, and V. S. Roitberg, {Sov. Phys. Usp.} {\bf 17}, 673 (1975).
\bibitem{Maimistov} A. I. Maimistov, A. M. Basharov, S. O. Elyutin, and Yu. M. Sklyarov, {Phys. Rep.} {\bf191}, 1 (1990).
\bibitem{Bonifacio} R. Bonifacio and L. A. Lugiato, {\pra} {\bf18}, 1129 (1978).
\bibitem{Hopf} F. A. Hopf, C. M. Bowden, and W. H. Louisell, {\pra} {\bf29}, 2591 (1984).
\bibitem{Bowd93} C. M. Bowden and J. P. Dowling, {\pra} {\bf47}, 1247 (1993).
\bibitem{Cren92} M. E. Crenshaw, M. Scalora, and C. M. Bowden, {\prl} {\bf68}, 911 (1992).
\bibitem{Bowd91} C. M. Bowden, A. Postan, and R. Inguva, {\josab} {\bf8}, 1081 (1991).
\bibitem{Golubev} N. V. Golubev and A. I. Kuleff, {\pra} {\bf90}, 035401 (2014).
\bibitem{Afan02} A. A. Afanas'ev, R. A. Vlasov, O. K. Khasanov, T. V. Smirnova, and O. M. Fedorova, {\josab} {\bf19}, 911 (2002).
\bibitem{Shaw} M. J. Shaw and B. W. Shore, {\josab} {\bf8}, 1127 (1990).
\bibitem{Novit2011} D. V. Novitsky, {\pra} {\bf84}, 013817 (2011).
\bibitem{Kozhekin} A. Kozhekin and G. Kurizki, {\prl} {\bf74}, 5020 (1995).
\bibitem{Kurizki} G. Kurizki, D. Petrosyan, T. Opatrny, M. Blaauboer, and B. Malomed, {\josab} {\bf19}, 2066 (2002).
\bibitem{Lemeza} R. A. Vlasov and A. M. Lemeza, {\pra} {\bf84}, 023828 (2011).
\bibitem{Ziolkowski} R. W. Ziolkowski, J. M. Arnold, and D. M. Gogny, {\pra} {\bf52}, 3082 (1995).
\bibitem{Novit2012b} D. V. Novitsky, {\pra} {\bf86}, 043835 (2012).
\bibitem{Cai2013} X. Cai, J. Zhao, Z. Wang, and Q. Lin, {J. Phys. B} {\bf46}, 175602 (2013).
\bibitem{Leblond} H. Leblond, H. Triki, and D. Mihalache, {Rom. Rep. Phys.} {\bf65}, 925 (2013).
\bibitem{Frantzeskakis} D. J. Frantzeskakis, H. Leblond, and D. Mihalache, {Rom. J. Phys.} {\bf59}, 767 (2014).
\bibitem{Manassah} J. T. Manassah, {\ao} {\bf25}, 3980 (1986).
\bibitem{Jha} P. K. Jha and Yu. V. Rostovtsev, {\pra} {\bf82}, 015801 (2010).
\bibitem{Ibanez} S. Ib\'{a}\~{n}ez, A. Peralta Conde, D. Gu\'{e}ry-Odelin, and J. G. Muga, {\pra} {\bf84}, 013428 (2011).
\bibitem{Yao} H. Yao, Y. Niu, Y. Peng, and S. Gong, {Chin. Opt. Lett.} {\bf10}, 011901 (2012).
\bibitem{Xu} Q. Q. Xu, D. Z. Yao, X. N. Liu, Q. Zhou, and G. G. Xiong, {\pra} {\bf86}, 023853 (2012).
\bibitem{Song} X. Song, S. Gong, W. Yang, S. Jin, X. Feng, and Z. Xu, {\oc} {\bf236}, 151 (2004).
\bibitem{Astapenko} V. A. Astapenko and M. S. Romodanovskii, {Laser Phys.} {\bf19}, 969 (2009).
\bibitem{Cai2015} X. Cai, Z. Wang, J. Zhao, and Q. Lin, {\oc} {\bf342}, 90 (2015).
\bibitem{Novit2012a} D. V. Novitsky, {\pra} {\bf86}, 063835 (2012).
\bibitem{Novit2014} D. V. Novitsky, {J. Phys. B} {\bf47}, 095401 (2014).
\bibitem{Novit2009} D. V. Novitsky, {\pra} {\bf79}, 023828 (2009).
\bibitem{Novit2010} D. V. Novitsky, {\pra} {\bf82}, 015802 (2010).
\end{thebibliography}
\end{document}